\documentclass[reprint,amsmath,amssymb,aps,prl,nobibnotes,floatfix,superscriptaddress,parskip=never]{revtex4-2}
\usepackage{etoolbox}
\usepackage{graphicx}
\usepackage{xcolor}
\definecolor{UniBlue}{RGB}{46,48,146}
\usepackage[
    colorlinks=true,
    citecolor=UniBlue,
    linkcolor=UniBlue,
    urlcolor=UniBlue]{hyperref}
\usepackage{bm}

\makeatletter
\pretocmd{\NAT@open}{\begingroup\color{\@citecolor}}{}{}
\apptocmd{\NAT@close}{\endgroup}{}{}
\makeatother

\usepackage{hyphenat}       

\begin{document}
\title{Directed Swarm Assembly due to Mixed Misaligned Perception-Dependent Motility}
\author{Rodrigo Saavedra}
\author{Marisol Ripoll}
 \email{m.ripoll@fz-juelich.de}
 \affiliation{
Theoretical Physics of Living Matter,  Institute for Advanced Simulation,
Forschungszentrum J\"ulich, 52425 J\"ulich, Germany
}
\date{\today}

\begin{abstract}
A mixture of particles with perception-dependent motility and opposite misaligned visual perception shows to spontaneously self-organize into a self-propelling bean-shaped cluster. %
The two species initially rotate in opposite directions, which together with the steric interactions, makes them segregate into two main counter-rotating domains forming a cohesive and persistently propelling single cluster. %
Mixtures of particles with misaligned perception and discontinuous motility are therefore a promising pathway for the design of programmable active matter. %
\end{abstract}

\maketitle
Active matter refers to systems of agents with a prescribed velocity due to a local consumption of energy~\cite{ramaswamy2003active}. %
Interacting active agents can self-organize leading to spectacular collective behavior, e.g. at the macroscopic level in schools of fish~\cite{tunstrom2013collective}, flocks of birds~\cite{ballerini2008interaction}, and swarms of insects~\cite{attanasi2014collective}. %
Self-organization is also observed at the microscopic level in living organisms like bacteria~\cite{aranson2022bacterial}, and active biopolymers~\cite{koenderink2009active}. %
Inspired on biology, there is  a growing interest in artificially engineered systems of activated constituents, e.g. light-activated colloidal suspensions~\cite{vutukuri2020lightswitchable,gomez-solano2020transient,zhang2020reconfigurable}, driven magnetic colloids~\cite{bricard2015emergenta,kawai2020degenerate,junot2021collective}, or micro-robots~\cite{yu2018pattern,xie2019reconfigurable,dorigo2021swarm,hussein2023actuation,yu2023swarming,hou2023review}. %
The processing of environmental information plays an essential role in the emergence of multiple collective states, such as herds of sheep~\cite{gomez-nava2022intermittent}, or schools of fish~\cite{wang2022impact}. %
To understand the interactions that lead to the different flocking states is one of the current most significant challenges in active matter~\cite{gompper2020,abaurreavelasco2018collective}. %
Currently, this has led to the development of the so-called programmable active matter~\cite{wang2017dynamic,nava2020novel,ceron2023programmable}, which aims to engineer systems of active automatons with adaptable collective states. %

Quorum sensing is a key mechanism by which agents sense and react to environmental information. 
Here, some of the system constitutive agents perceive their local surroundings and then switch an intrinsic property, such as their individual motility, when a certain condition is fulfilled. %
As an example, certain populations of bacteria like \emph{Aliivibrio fischerei}~\cite{stevens1997quorum} fluoresce altogether after a chemical density threshold is surpassed. %
For synthetic microswimmers this has also been experimentally realized for externally controlled colloidal suspensions, where a narrow laser beam locally activates colloids according to a computer assisted feedback mechanism accounting for the local density perception, showing the particle aggregation into a circular cluster~\cite{bauerle2018selforganization}. %
Quorum sensing considers therefore an isotropic perception of the agents. 
Non-reciprocity can be also taken into account by the restriction of the perception to that occurring inside a well-defined vision cone. 
This has shown to lead to a plethora of collective behaviors that do not occur when interactions are reciprocal. %
For example, active particles with visual perception aligning their orientation based on the position of other neighbouring particles, have shown to aggregate into clusters, polar filaments, and nematic bands~\cite{barberis2016largescale,costanzo2019millinginduction,negi2022emergent,negi2024collective}, as has been demonstrated in particle-based numerical simulations. %
In experiments, visual perception has also shown to generate vortical structures when imposing a local torque to light-activated colloids~\cite{bauerle2020formation}. %

A different framework in which flocking has shown to emerge is for cases combining motility with another local interaction, such as steric forces or aligning torques. %
In recent years, mixtures of particles displaying motility-induced self-organization into directed swarms have been investigated for mixtures of active and passive Brownian particles forming propagating interfaces~\cite{wysocki2016propagating}, for colloids with non-reciprocal phoretic interactions aggregating into a motile species-separated structure resembling a macroscopic Janus colloid~\cite{agudo-canalejo2019active}, as well as for light-activated colloids with aligning interactions that form an arrow-like swarm escaping from a predator~\cite{chen2022collective}. %
In single-species systems, motility-induced flocking has been found in systems of self-propelled rods~\cite{bar2020selfpropelled}, as well as in systems of active particles with attractions~\cite{caprini2023flocking}. %

In this letter, we propose an experimentally feasible mechanism by which particles sedimented on a surface self-assemble into a single swarming cluster. %
Two types of active particles with perception displaced left and right from the motion directions are considered. %
Each particle type initially shows an effective rotation around the cluster center with clock or anti-clockwise directions depending on their misalignment. %
These rotations, together with the effect of steric interactions, trigger the assembly of a cohesive bean-shaped cluster with two counter-rotating lobes, each rich in one of the particle types, leading to the persistent motion of the cluster. %
We describe the procedure that leads to the emergence of this phenomena considering the individual particle behavior and then characterize it in terms of the misalignment angle.

\begin{figure}[h!]
    \centering
    \includegraphics[width=\linewidth]{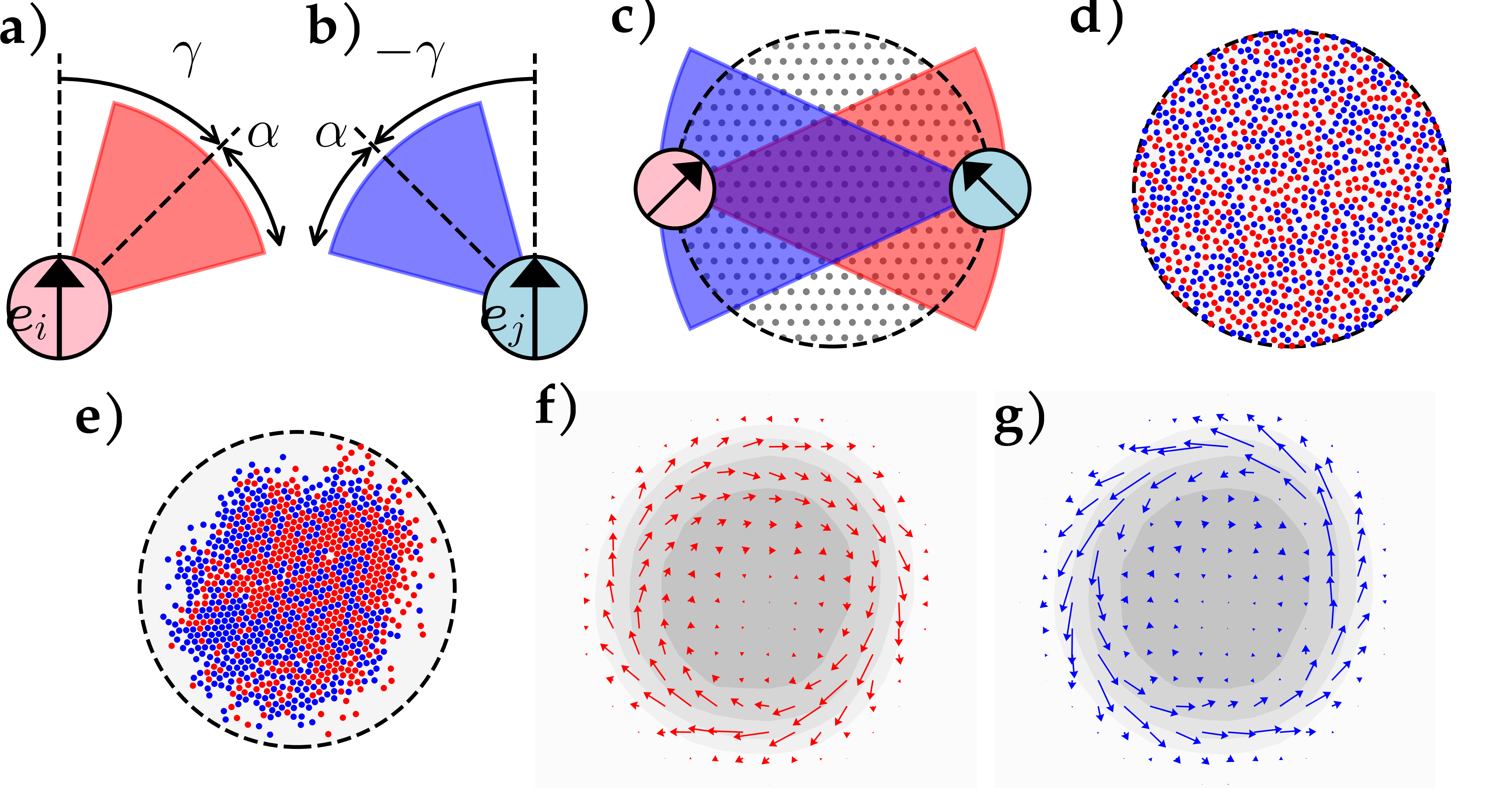}
    \caption{\label{fig:intro}%
    Sketches of active particles with oppositely misaligned visual perceptions. Propulsion occurs in the $\bm{e}_i$, $\bm{e}_j$ directions. 
    The vision cones half-width-angle is~$\alpha$, and their symmetry axes have a relative angle $\gamma$ with the propulsion directions, this is the misalignment, which is oriented towards the right for the red particles in a) and towards the left for the blue ones in b). 
    c)~Sketch of particles in a cluster boundary with maximum perception.  
    d)~Initial homogeneous configuration of an equimolar mixture of particles with perception-dependent motility with oppositely misaligned perceptions. 
    e)~Initial collapsed configuration after very short time. 
    f),~g)~Average displacement of each particle type in the collapsed state showing the early counter-rotation of particles with right-handed misalignment in f) and left-handed in g).    
    }
\end{figure}

The system is composed 
of $N$ particles characterized by their position $\bm{r}_i$,  and propulsion orientation, $\bm{e}_i\equiv(\cos\phi_i,\sin\phi_i)^T$,  with $\phi_i$ the angle between the orientation and the $x$ axis. %
Each particle $i$ perceives all neighbouring particles $j$ placed inside the perception cone, via the function, $P_i = \sum_{j} (1/r_{ij}) \Theta(r_{ij}- r_c) $, with $r_{ij}$ the interparticle distance and  $r_c$ the maximum perception distance, similar to previous works~\cite{lavergne2019group,saavedra2024swirling}. %
The cone half-width angle is $\alpha$, while $\gamma$ is the angle between the cone's symmetry axis and the particle self-propulsion direction $\bm{e}_i$, as illustrated in Fig.~\ref{fig:intro}a. %
The particles dynamics is determined by overdamped Brownian dynamics, %
\begin{equation}
    \begin{aligned}
        \dot{\bm{r}}_i &= v_i\bm{e}_i + \bm{f}^\text{EV}_i+ \sqrt{2 D_t}\bm{\xi}_i, \\
        \dot{\phi}_i &= \sqrt{2 D_r} \eta_i,
    \end{aligned}%
    \label{eqs:overdamped_langevin}%
 \end{equation}%
where, $\bm{\xi}_i$ and $\eta_i$ are translational and orientational white noises, and $D_t$ and $D_r$ the translational and rotational diffusion coefficients. %
The excluded volume force is~\mbox{$\bm{f}^\text{EV}=-\nabla{U}$}, with $U(\sigma,\epsilon)$ the Weeks-Chandler-Anderson potential, with $\sigma$ the particle diameter, and $\epsilon=100k_\mathrm{B}T$ the repulsion strength. %
The perception dependent particle velocity is defined by~$v_i=v_0\Theta(q_i - q^*)$, with~$v_0$ a constant self-propulsion velocity, and~$q^*=P^*_i/P_0$ with $P^*_i$ the threshold perception value, and a constant normalization factor $P_0=\alpha \rho_0 R_0$ dependent on the system parameters, $\rho_0=N/(\pi R_0^2)$ the system number density, and $R_0=2r_c$ also typically the initial cluster radius. %
This means that only particles with a large enough number of neighbors in their vision cone will become active. %
This mechanism induces cluster cohesion, which collectively rotate when the propulsion direction and cone direction are misalignment, this is for $\gamma \neq 0$~\cite{saavedra2024swirling}. %
The Euler algorithm is used to integrate Eq.~(\ref{eqs:overdamped_langevin}) with $\Delta t=10^{-5}$, and all quantities are normalized or expressed in simulation units, here $\sigma$ and $D_t$. %
Systems with $N=1000$ particles are here studied with $v_0=40$, $D_r= 8.3$, $\rho_0=0.51$, $k_\mathrm{B}T=1$, and $\alpha=\pi/4$, which correspond to a P{\'e}clet number $Pe=4.8$,  with $Pe = v_0/(\sigma D_r)$. We use $\gamma=\pi/4$ unless otherwise specified. 

An equimolar mixture of particles is here considered, which are identical except the misalignment of their vision cones, see Fig.~\ref{fig:intro}a,b. %
This means that to become active particles each of species need to have opposite orientation relative to the cluster center, see Fig.~\ref{fig:intro}c. %
Simulations start from an homogeneous distribution in a circular area of radius $R_0$, as shown in Fig.~\ref{fig:intro}d. %
At first, particles with some orientation towards the cluster center become active, such that the system shows a fast initial collapse into a compact homogeneous structure as shown in Fig.~\ref{fig:intro}e. %
In this configuration, particles in the cluster center are mainly active due to their large perception, although their actual motion is blocked due to steric interactions. %
Density in the external cluster layer is lower and colloids become active only if they are co-oriented with their intrinsic misalignment. %
This means that right-misaligned particles tend to rotate around the whole cluster with a clock-wise direction, and conversely left-misaligned particles, as clearly shown in Fig.~\ref{fig:intro}f,g. 

\begin{figure}[ht]
    \centering
    \includegraphics[width=\linewidth]{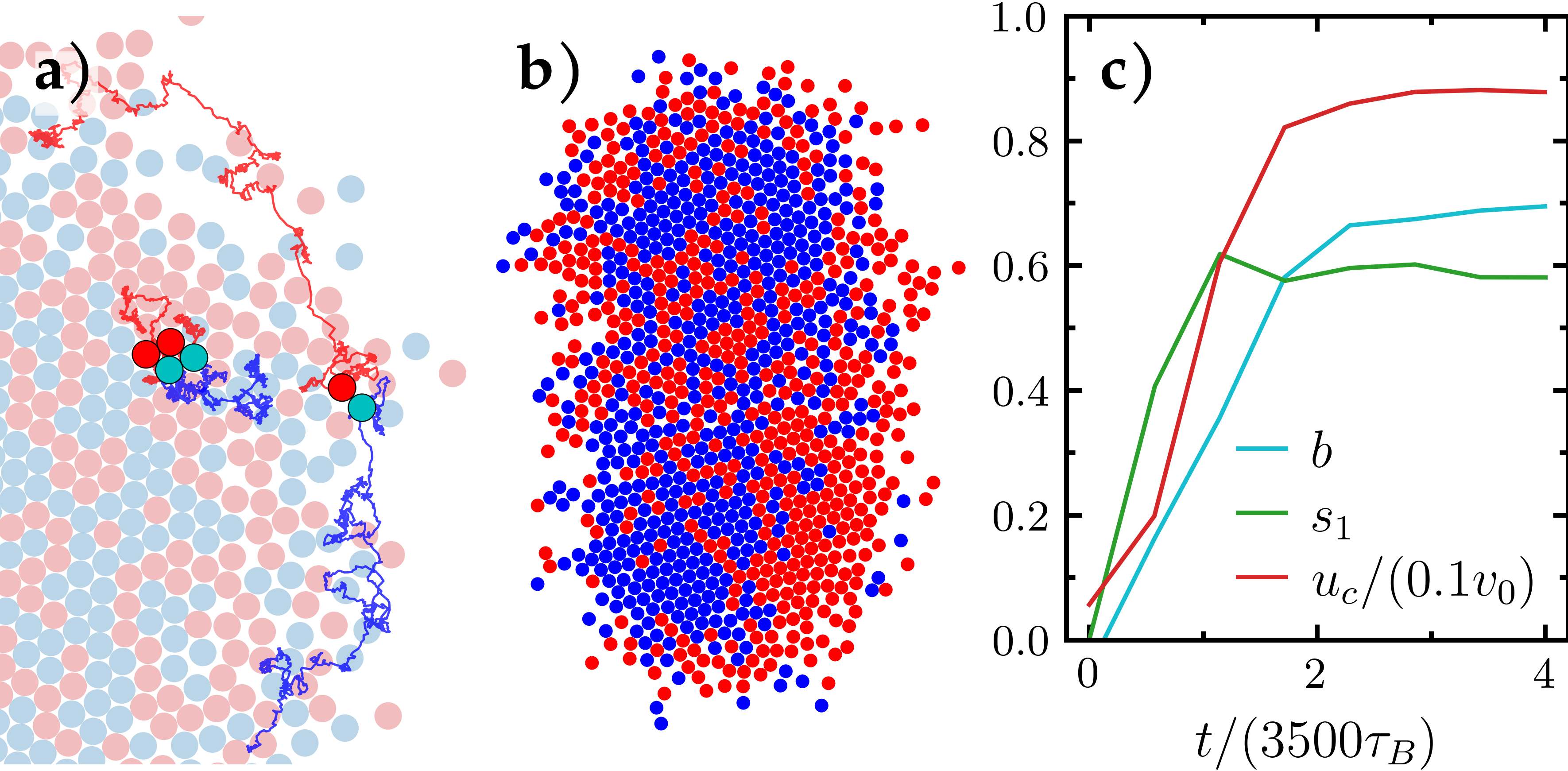}
    \caption{\label{fig:separation} %
    Early separation snapshot where small red and blue domains start to differentiate: 
    a)~partial snapshots with trajectories of a few tagged particles, illustrating particle collision and initial self-sorting procedure;
    b)~complete cluster showing the initial elongation. %
    c)~Time evolution of the separation parameter $b$, the polar order of the active particles $s_1$ and the cluster center of mass velocity $u_c$. %
    The initial homogeneous, non-polar and static  configuration shows to reach a steady separated, actively polar, and self-propelled conformation. Time is normalized by half-separation time which shows to be $3500\tau_B$, with $\tau_B=\sigma/v_0$. %
  }
\end{figure}

Particles in the external cluster layer change their properties by rotational diffusion or by colliding with other particles, specially counter-rotating particles. 
When the new position and orientation perceive not enough neighbors,  %
the particle remains passive until the next reorientation, which  occurs when the particle has not moved significantly far away from the cluster. %
When the new position and orientation perceive enough neighbors, 
the particle might move together with another one of the same type, away from a particle of those of the other type, or get stalled in a jammed configuration, see Fig.~\ref{fig:separation}a. %
On average, this already favors segregation by particle-type. %
In this way, small domains form and grow within the cluster where one particle type is more frequent than the other. %
The domain formation on the cluster edge perturbs the rotation of particles of opposite type. %
This spontaneously breaks down the spatial cluster symmetry and results into an elongation of the cluster, as can be seen in Fig.~\ref{fig:separation}b. %
With the increase of the average domains size, the average cluster alignment produces also larger rearrangements in neighboring domains reaching the cluster center, with the eventual formation of short-lived dislocations, which enhance the domains rearrangement and growth, see movie in SM~\cite{sm}. %
When the domains are large enough they start to rotate in the direction of the dominant particle type. %

In order to quantitatively characterize the self-sorting procedure, we consider three quantities. %
First, we define the separation parameter~$b$, as a useful alternative to the standard Moran's index~\cite{moran,fischer23}. %
To compute this parameter, we consider as neighboring particles those that are separated by a small distance of~$r=2\sigma$. %
For each particle~$j$, we count the number of neighbors of the same type~$n_{j,s}$ (these are left for a left particle, or conversely right for a right particle), the number of neighbors of the opposite type~$n_{j,s}$ (right for a left particle), and we average and normalize by the total number of neighbors as %
\begin{equation}
b = \frac{1}{N}\sum_{j=1}^N\frac{n_{j,s}-n_{j,o}}{n_{j,s}+n_{j,o}}. 
\end{equation}
With this definition, the separation parameter varies from~$b=0$ for perfectly homogeneous configurations, to~$b=1$ for perfect separation without any interface, such that in between values are obtained when one particle type rich subdomains form. %
The time increase of~$b$ displayed in~Fig.~\ref{fig:separation}c, shows an initially homogeneous configuration which progressively separates, reaching a steady state value at longer times. %
At long times, the separation value saturates in a steady state value, which for the standard model parameters here employed is~$b \simeq 0.7$. 
Time is normalized with the time at which the cluster velocity has reached half the saturation value, which for the standard parameters here used is roughly~$\tau_s\simeq 3500 \tau_B$, with $\tau_B=\sigma/v_0$ the single self-propelled particle ballistic time. 

We also compute the overall polar order of the particles displacements $s_1$, defined as %
\begin{equation}
    s_1(t) = \bigg|\frac{1}{N}\sum_{i=1}^{N}e^{i\theta_i(t)}\bigg|,
\end{equation}
where~$\theta_i$ corresponds to the polar angle of the particle normalized effective velocity~$\hat{\bm{v}}_i=\bm{v}_i/v_i$, and~$\bm{v}_i=\Delta{\bm{r}}_i/\Delta{t}$ is calculated from the particle displacement in the centre of mass reference frame during a fixed time interval, here~$\Delta{t}=20\tau_B$. %
Polar order increases similar to $b$ saturating to a persistent value, as shown in Fig.~\ref{fig:separation}c. %
Similarly to the case of single component clusters~\cite{saavedra2024swirling}, there is no overall alignment of the particles actual orientation since the polar order relates to the particle displacements, which occurs all in the same direction. %

The large polarity of the active particles has as a consequence the overall motion of the cluster. %
Assuming a ballistic self-propulsion, the displacement of the cluster-center-of mass is averaged~$\Delta{t}=400\tau_B$ and then averaged over $10$ realizations, such that a time dependent velocity $u_c$ is computed, see Fig.~\ref{fig:separation}c. %
This averaged velocity starts from a vanishing value in the initial non-separated state, and increases until a constant velocity which, for our parameters, is $10\%$ of the velocity of a single active colloid, $v_0$. %
Separation parameter, induced polarity and cluster velocity are therefore strongly interconnected. %
The time evolution of the three quantities towards the steady state is not completely simultaneous, as shown in Fig.~\ref{fig:separation}c. %
Polarity is the first quantity that grows and reaches an steady state value, closely followed by the velocity of the cluster center of mass. %
The separation parameter shows a persistent increase from the initial disorder state, but takes slightly longer to fully develop. %

\begin{figure}[h!]
    \centering
    \includegraphics[width=\linewidth]{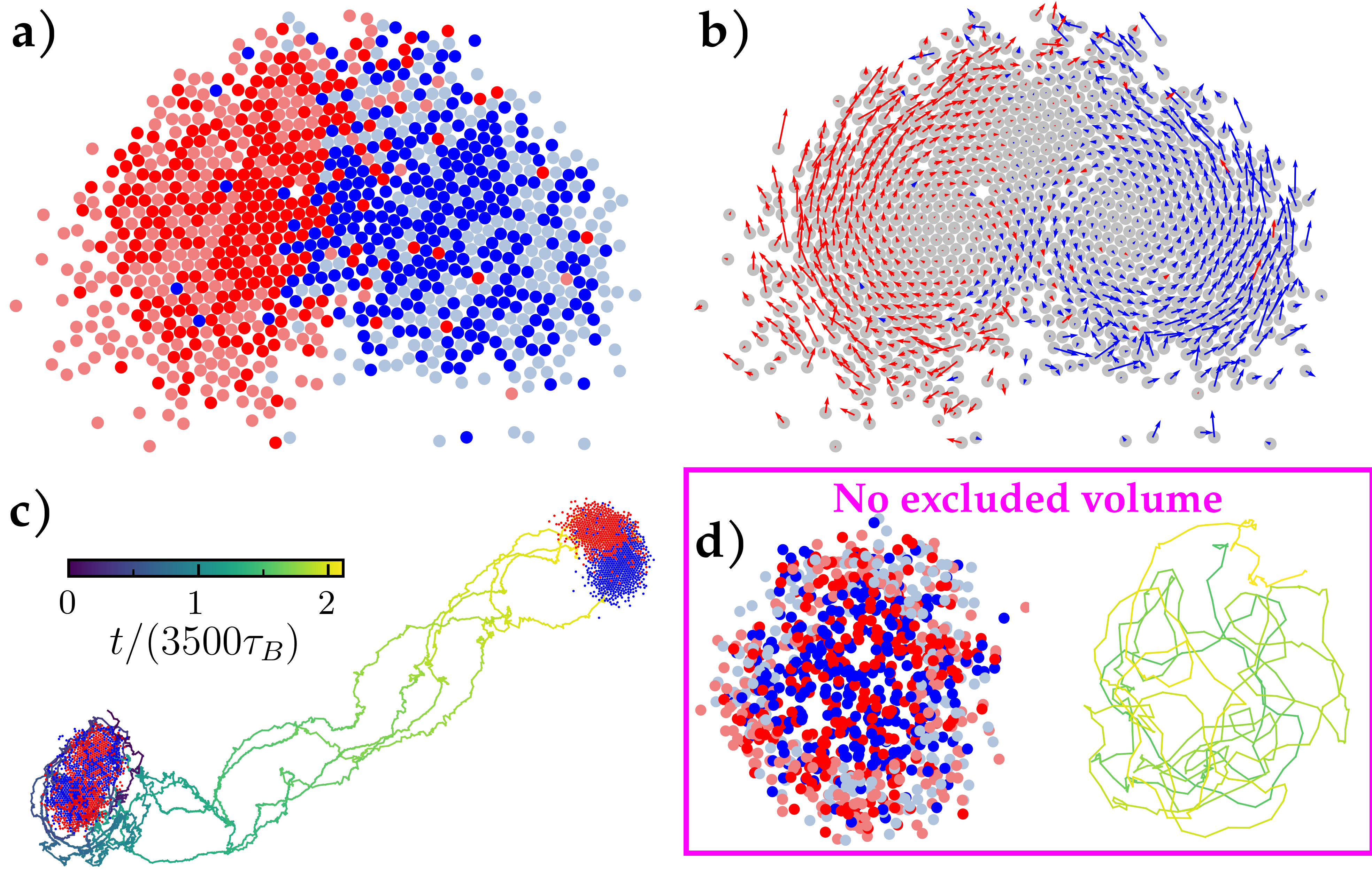}
    \caption{\label{fig:propulsion} %
    Snapshots of the cluster in a steady state, where the two counter-rotating domains can be distinguished, in 
    a)~full (shaded) colors correspond active (passive) particles, in b)~arrows depict the direction and speed of the particle displacements, in the cluster center-of-mass reference frame. 
    c)~Trajectories of a few particles within the cluster, showing the overall cluster displacement. Color bar indicates time evolution, and the small snapshots illustrate the corresponding cluster states. 
    d)~Snapshot and trajectories of a cluster in which particles have no excluded volume interactions, showing that these are crucial for the described particle segregation and overall cluster self-propulsion. 
     }
\end{figure}
When the maximum separation is reached, the cluster organizes in two main domains of typically the same size. %
Each domain is composed by mainly one particle type, as shown in Fig.~\ref{fig:propulsion}a, and each of them rotates according to the misalignment of the majority, as shown in Fig.~\ref{fig:propulsion}b. %
The high perception in the cluster center makes that many more particles are active than in the outside of the cluster. %
The two counter-rotating domains remain completely adhered to each other, and given the difference in density, the particles at the center move much slower than those in the external cluster layer. %
This also  makes that the two stagnation points are not central to the respective domains, but are placed closer to the cluster center, see Fig.~\ref{fig:propulsion}b. %
Active particles at the outside part of the cluster drag all other particles along the corresponding rotation, regardless if the are active or passive, or even of the opposite particle type. %
The cluster polar order, self-propulsion direction and overall cluster motion are therefore determined by the few active particles placed at the outside cluster shell, which also deforms the cluster shape from elongated to a bean shape with a tip at the motion direction. 
The trajectory of a few tagged particles is presented in Fig.~\ref{fig:propulsion}c in the laboratory reference frame, see also movie in SM~\cite{sm}. %
The displayed motion shows to be clearly periodic showing how each particle rotates within one of the domains and how the cluster propels in a well defined direction for a long period of time.  

As verification of the presented mechanism, we perform a few simulations in which the excluded volume between particles is switched off. %
An snapshot of a late conformation and one particle trajectory is presented in Fig.~\ref{fig:propulsion}d clearly showing  that there is no separation into domains of different particle types and also no cluster self-propulsion. %
The presence of steric interactions is therefore crucial for the self-assembly of the directed swarm.

\begin{figure}[h!]
    \centering
    \includegraphics[width=\linewidth]{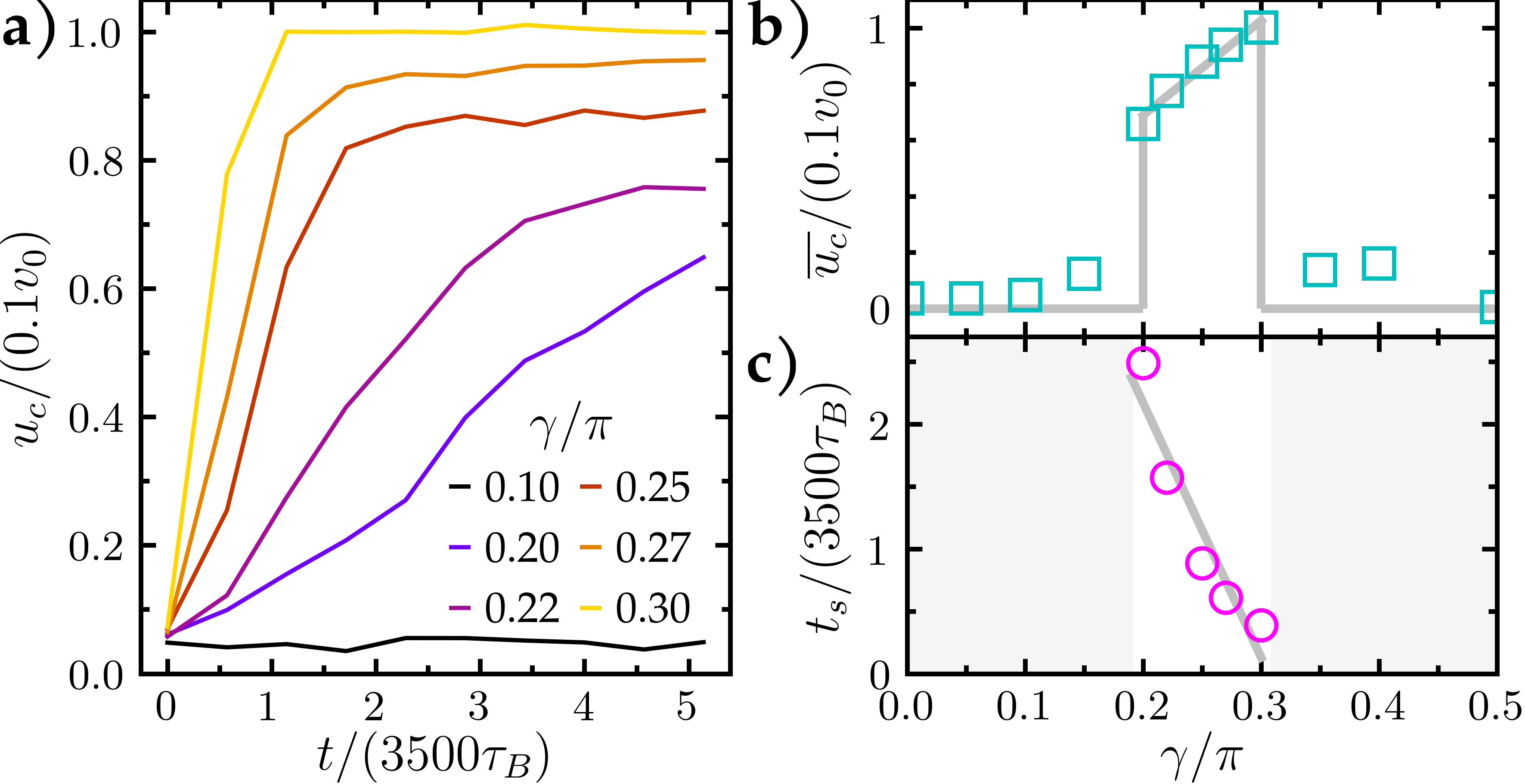}
    \caption{\label{fig:analysis}%
     a)~Time evolution of the cluster center of mass velocity for systems with varying misalignment angle~$\gamma$ where the system. %
     b)~Steady state values of the cluster velocity, and %
     c)~characteristic time to reach the steady steady state. %
     Grey lines in b, c are a guide to eye. 
     }
\end{figure}
Our discussion until now has just considered a fixed set of parameters, which result into a stable solid behavior, as well as a particular value of the steady state value of the cluster velocity~$u_c$, motion persistence, or characteristic self-assembly times. %
Various parameters are expected to have a significant influence, such as the width of the vision cone~$\alpha$, the perception range and threshold or the number of considered particles. %
Here we first concentrate on the influence of the misalignment angle~$\gamma$. %
Since the sorting mechanism is based on the steric interactions between particles with opposite directions of misalignment, it is clear that very small values of~$\gamma$ are not expected to induce any swirling, and therefore also not separation effects. %
On the other limit, a misalignment close to~$\pi/2$ is not able to keep the cohesion of the cluster in the case of single component clusters~\cite{saavedra2024swirling}. %
Results in Fig.~\ref{fig:analysis}a show the increase of the steady state velocity with misalignment, as well as a speed up of the procedure. %
In Fig.~\ref{fig:analysis}b, the measured steady-state velocities $\bar{u}_c$, show to be non-vanishing only in the interval $0.2 \leq \gamma/\pi \leq 0.3$, and also that within this interval the increase of $\bar{u}_c$ is approximately linear. %
The measured characteristic self-assembly times are shown in Fig.~\ref{fig:analysis}c, conversely show a linearly decrease with $\gamma$, within the same interval as the $\bar{u}_c$. 
For misalignment outside this range, swirling is observed in single component systems but the self-sorting procedure is not strong enough in comparison to the related fluctuations to result into the assembly of a coherent swarming cluster.

In conclusion, an initially homogeneous mixture of particles with oppositely misaligned perception-dependent motility show to undergo a self-organized sorting procedure resulting into a cohesive and persistently propagating cluster. The resulting cluster is elongated and organized in two main counter-rotating subdomains. 
The overall motion is then driven by the overall alignment of the active particles velocities. 
The mechanism here described is completely different to previous studies, since due to the misaligned perception dependent motility, it does not require of any explicit particle alignment, nor external torques, or forces,  it being still able to result into a motion in a well defined linear momentum.   
The swarming strategy can be implemented in different types of experiments such as colloids activated by light or in robot ensembles, serving then various potential purposes.

\begin{acknowledgements}
This work was financially supported by the CONACyT-DAAD scholarship program. %
The authors gratefully acknowledge the computing time granted by the JARA Vergabegremium and provided on the JARA Partition part of the supercomputer JURECA at Forschungszentrum J\"ulich~\cite{jureca2018}.
\end{acknowledgements}
  

\begin{thebibliography}{43}%
\makeatletter
\providecommand \@ifxundefined [1]{%
 \@ifx{#1\undefined}
}%
\providecommand \@ifnum [1]{%
 \ifnum #1\expandafter \@firstoftwo
 \else \expandafter \@secondoftwo
 \fi
}%
\providecommand \@ifx [1]{%
 \ifx #1\expandafter \@firstoftwo
 \else \expandafter \@secondoftwo
 \fi
}%
\providecommand \natexlab [1]{#1}%
\providecommand \enquote  [1]{``#1''}%
\providecommand \bibnamefont  [1]{#1}%
\providecommand \bibfnamefont [1]{#1}%
\providecommand \citenamefont [1]{#1}%
\providecommand \href@noop [0]{\@secondoftwo}%
\providecommand \href [0]{\begingroup \@sanitize@url \@href}%
\providecommand \@href[1]{\@@startlink{#1}\@@href}%
\providecommand \@@href[1]{\endgroup#1\@@endlink}%
\providecommand \@sanitize@url [0]{\catcode `\\12\catcode `\$12\catcode
  `\&12\catcode `\#12\catcode `\^12\catcode `\_12\catcode `\%12\relax}%
\providecommand \@@startlink[1]{}%
\providecommand \@@endlink[0]{}%
\providecommand \url  [0]{\begingroup\@sanitize@url \@url }%
\providecommand \@url [1]{\endgroup\@href {#1}{\urlprefix }}%
\providecommand \urlprefix  [0]{URL }%
\providecommand \Eprint [0]{\href }%
\providecommand \doibase [0]{https://doi.org/}%
\providecommand \selectlanguage [0]{\@gobble}%
\providecommand \bibinfo  [0]{\@secondoftwo}%
\providecommand \bibfield  [0]{\@secondoftwo}%
\providecommand \translation [1]{[#1]}%
\providecommand \BibitemOpen [0]{}%
\providecommand \bibitemStop [0]{}%
\providecommand \bibitemNoStop [0]{.\EOS\space}%
\providecommand \EOS [0]{\spacefactor3000\relax}%
\providecommand \BibitemShut  [1]{\csname bibitem#1\endcsname}%
\let\auto@bib@innerbib\@empty
\bibitem [{\citenamefont {Ramaswamy}\ \emph {et~al.}(2003)\citenamefont
  {Ramaswamy}, \citenamefont {Simha},\ and\ \citenamefont
  {Toner}}]{ramaswamy2003active}%
  \BibitemOpen
  \bibfield  {author} {\bibinfo {author} {\bibfnamefont {S.}~\bibnamefont
  {Ramaswamy}}, \bibinfo {author} {\bibfnamefont {R.~A.}\ \bibnamefont
  {Simha}},\ and\ \bibinfo {author} {\bibfnamefont {J.}~\bibnamefont {Toner}},\
  }\href {https://doi.org/10.1209/epl/i2003-00346-7} {\bibfield  {journal}
  {\bibinfo  {journal} {EPL}\ }\textbf {\bibinfo {volume} {62}},\ \bibinfo
  {pages} {196} (\bibinfo {year} {2003})}\BibitemShut {NoStop}%
\bibitem [{\citenamefont {Tunstr{\o}m}\ \emph {et~al.}(2013)\citenamefont
  {Tunstr{\o}m}, \citenamefont {Katz}, \citenamefont {Ioannou}, \citenamefont
  {Huepe}, \citenamefont {Lutz},\ and\ \citenamefont
  {Couzin}}]{tunstrom2013collective}%
  \BibitemOpen
  \bibfield  {author} {\bibinfo {author} {\bibfnamefont {K.}~\bibnamefont
  {Tunstr{\o}m}}, \bibinfo {author} {\bibfnamefont {Y.}~\bibnamefont {Katz}},
  \bibinfo {author} {\bibfnamefont {C.~C.}\ \bibnamefont {Ioannou}}, \bibinfo
  {author} {\bibfnamefont {C.}~\bibnamefont {Huepe}}, \bibinfo {author}
  {\bibfnamefont {M.~J.}\ \bibnamefont {Lutz}},\ and\ \bibinfo {author}
  {\bibfnamefont {I.~D.}\ \bibnamefont {Couzin}},\ }\href
  {https://doi.org/10.1371/journal.pcbi.1002915} {\bibfield  {journal}
  {\bibinfo  {journal} {PLoS Comput. Biol.}\ }\textbf {\bibinfo {volume} {9}},\
  \bibinfo {pages} {e1002915} (\bibinfo {year} {2013})}\BibitemShut {NoStop}%
\bibitem [{\citenamefont {Ballerini}\ \emph {et~al.}(2008)\citenamefont
  {Ballerini}, \citenamefont {Cabibbo}, \citenamefont {Candelier},
  \citenamefont {Cavagna}, \citenamefont {Cisbani}, \citenamefont {Giardina},
  \citenamefont {Lecomte}, \citenamefont {Orlandi}, \citenamefont {Parisi},
  \citenamefont {Procaccini}, \citenamefont {Viale},\ and\ \citenamefont
  {Zdravkovic}}]{ballerini2008interaction}%
  \BibitemOpen
  \bibfield  {author} {\bibinfo {author} {\bibfnamefont {M.}~\bibnamefont
  {Ballerini}}, \bibinfo {author} {\bibfnamefont {N.}~\bibnamefont {Cabibbo}},
  \bibinfo {author} {\bibfnamefont {R.}~\bibnamefont {Candelier}}, \bibinfo
  {author} {\bibfnamefont {A.}~\bibnamefont {Cavagna}}, \bibinfo {author}
  {\bibfnamefont {E.}~\bibnamefont {Cisbani}}, \bibinfo {author} {\bibfnamefont
  {I.}~\bibnamefont {Giardina}}, \bibinfo {author} {\bibfnamefont
  {V.}~\bibnamefont {Lecomte}}, \bibinfo {author} {\bibfnamefont
  {A.}~\bibnamefont {Orlandi}}, \bibinfo {author} {\bibfnamefont
  {G.}~\bibnamefont {Parisi}}, \bibinfo {author} {\bibfnamefont
  {A.}~\bibnamefont {Procaccini}}, \bibinfo {author} {\bibfnamefont
  {M.}~\bibnamefont {Viale}},\ and\ \bibinfo {author} {\bibfnamefont
  {V.}~\bibnamefont {Zdravkovic}},\ }\href
  {https://doi.org/10.1073/pnas.0711437105} {\bibfield  {journal} {\bibinfo
  {journal} {PNAS}\ }\textbf {\bibinfo {volume} {105}},\ \bibinfo {pages}
  {1232} (\bibinfo {year} {2008})}\BibitemShut {NoStop}%
\bibitem [{\citenamefont {Attanasi}\ \emph {et~al.}(2014)\citenamefont
  {Attanasi}, \citenamefont {Cavagna}, \citenamefont {Castello}, \citenamefont
  {Giardina}, \citenamefont {Melillo}, \citenamefont {Parisi}, \citenamefont
  {Pohl}, \citenamefont {Rossaro}, \citenamefont {Shen}, \citenamefont
  {Silvestri},\ and\ \citenamefont {Viale}}]{attanasi2014collective}%
  \BibitemOpen
  \bibfield  {author} {\bibinfo {author} {\bibfnamefont {A.}~\bibnamefont
  {Attanasi}}, \bibinfo {author} {\bibfnamefont {A.}~\bibnamefont {Cavagna}},
  \bibinfo {author} {\bibfnamefont {L.~D.}\ \bibnamefont {Castello}}, \bibinfo
  {author} {\bibfnamefont {I.}~\bibnamefont {Giardina}}, \bibinfo {author}
  {\bibfnamefont {S.}~\bibnamefont {Melillo}}, \bibinfo {author} {\bibfnamefont
  {L.}~\bibnamefont {Parisi}}, \bibinfo {author} {\bibfnamefont
  {O.}~\bibnamefont {Pohl}}, \bibinfo {author} {\bibfnamefont {B.}~\bibnamefont
  {Rossaro}}, \bibinfo {author} {\bibfnamefont {E.}~\bibnamefont {Shen}},
  \bibinfo {author} {\bibfnamefont {E.}~\bibnamefont {Silvestri}},\ and\
  \bibinfo {author} {\bibfnamefont {M.}~\bibnamefont {Viale}},\ }\href
  {https://doi.org/10.1371/journal.pcbi.1003697} {\bibfield  {journal}
  {\bibinfo  {journal} {PLOS Comput. Biol.}\ }\textbf {\bibinfo {volume}
  {10}},\ \bibinfo {pages} {e1003697} (\bibinfo {year} {2014})}\BibitemShut
  {NoStop}%
\bibitem [{\citenamefont {Aranson}(2022)}]{aranson2022bacterial}%
  \BibitemOpen
  \bibfield  {author} {\bibinfo {author} {\bibfnamefont {I.~S.}\ \bibnamefont
  {Aranson}},\ }\href {https://doi.org/10.1088/1361-6633/ac723d} {\bibfield
  {journal} {\bibinfo  {journal} {Rep. Prog. Phys.}\ }\textbf {\bibinfo
  {volume} {85}},\ \bibinfo {pages} {076601} (\bibinfo {year}
  {2022})}\BibitemShut {NoStop}%
\bibitem [{\citenamefont {Koenderink}\ \emph {et~al.}(2009)\citenamefont
  {Koenderink}, \citenamefont {Dogic}, \citenamefont {Nakamura}, \citenamefont
  {Bendix}, \citenamefont {MacKintosh}, \citenamefont {Hartwig}, \citenamefont
  {Stossel},\ and\ \citenamefont {Weitz}}]{koenderink2009active}%
  \BibitemOpen
  \bibfield  {author} {\bibinfo {author} {\bibfnamefont {G.~H.}\ \bibnamefont
  {Koenderink}}, \bibinfo {author} {\bibfnamefont {Z.}~\bibnamefont {Dogic}},
  \bibinfo {author} {\bibfnamefont {F.}~\bibnamefont {Nakamura}}, \bibinfo
  {author} {\bibfnamefont {P.~M.}\ \bibnamefont {Bendix}}, \bibinfo {author}
  {\bibfnamefont {F.~C.}\ \bibnamefont {MacKintosh}}, \bibinfo {author}
  {\bibfnamefont {J.~H.}\ \bibnamefont {Hartwig}}, \bibinfo {author}
  {\bibfnamefont {T.~P.}\ \bibnamefont {Stossel}},\ and\ \bibinfo {author}
  {\bibfnamefont {D.~A.}\ \bibnamefont {Weitz}},\ }\href@noop {} {\bibfield
  {journal} {\bibinfo  {journal} {PNAS}\ }\textbf {\bibinfo {volume} {106}},\
  \bibinfo {pages} {15192} (\bibinfo {year} {2009})}\BibitemShut {NoStop}%
\bibitem [{\citenamefont {Vutukuri}\ \emph {et~al.}(2020)\citenamefont
  {Vutukuri}, \citenamefont {Lisicki}, \citenamefont {Lauga},\ and\
  \citenamefont {Vermant}}]{vutukuri2020lightswitchable}%
  \BibitemOpen
  \bibfield  {author} {\bibinfo {author} {\bibfnamefont {H.~R.}\ \bibnamefont
  {Vutukuri}}, \bibinfo {author} {\bibfnamefont {M.}~\bibnamefont {Lisicki}},
  \bibinfo {author} {\bibfnamefont {E.}~\bibnamefont {Lauga}},\ and\ \bibinfo
  {author} {\bibfnamefont {J.}~\bibnamefont {Vermant}},\ }\href
  {https://doi.org/10.1038/s41467-020-15764-1} {\bibfield  {journal} {\bibinfo
  {journal} {Nat. Commun.}\ }\textbf {\bibinfo {volume} {11}},\ \bibinfo
  {pages} {2628} (\bibinfo {year} {2020})}\BibitemShut {NoStop}%
\bibitem [{\citenamefont {{Gomez-Solano}}\ \emph {et~al.}(2020)\citenamefont
  {{Gomez-Solano}}, \citenamefont {Roy}, \citenamefont {Araki}, \citenamefont
  {Dietrich},\ and\ \citenamefont {Macio{\l}ek}}]{gomez-solano2020transient}%
  \BibitemOpen
  \bibfield  {author} {\bibinfo {author} {\bibfnamefont {J.~R.}\ \bibnamefont
  {{Gomez-Solano}}}, \bibinfo {author} {\bibfnamefont {S.}~\bibnamefont {Roy}},
  \bibinfo {author} {\bibfnamefont {T.}~\bibnamefont {Araki}}, \bibinfo
  {author} {\bibfnamefont {S.}~\bibnamefont {Dietrich}},\ and\ \bibinfo
  {author} {\bibfnamefont {A.}~\bibnamefont {Macio{\l}ek}},\ }\href
  {https://doi.org/10.1039/D0SM00964D} {\bibfield  {journal} {\bibinfo
  {journal} {Soft Matter}\ }\textbf {\bibinfo {volume} {16}},\ \bibinfo {pages}
  {8359} (\bibinfo {year} {2020})}\BibitemShut {NoStop}%
\bibitem [{\citenamefont {Zhang}\ \emph {et~al.}(2020)\citenamefont {Zhang},
  \citenamefont {Sokolov},\ and\ \citenamefont
  {Snezhko}}]{zhang2020reconfigurable}%
  \BibitemOpen
  \bibfield  {author} {\bibinfo {author} {\bibfnamefont {B.}~\bibnamefont
  {Zhang}}, \bibinfo {author} {\bibfnamefont {A.}~\bibnamefont {Sokolov}},\
  and\ \bibinfo {author} {\bibfnamefont {A.}~\bibnamefont {Snezhko}},\ }\href
  {https://doi.org/10.1038/s41467-020-18209-x} {\bibfield  {journal} {\bibinfo
  {journal} {Nat. Commun.}\ }\textbf {\bibinfo {volume} {11}},\ \bibinfo
  {pages} {4401} (\bibinfo {year} {2020})}\BibitemShut {NoStop}%
\bibitem [{\citenamefont {Bricard}\ \emph {et~al.}(2015)\citenamefont
  {Bricard}, \citenamefont {Caussin}, \citenamefont {Das}, \citenamefont
  {Savoie}, \citenamefont {Chikkadi}, \citenamefont {Shitara}, \citenamefont
  {Chepizhko}, \citenamefont {Peruani}, \citenamefont {Saintillan},\ and\
  \citenamefont {Bartolo}}]{bricard2015emergenta}%
  \BibitemOpen
  \bibfield  {author} {\bibinfo {author} {\bibfnamefont {A.}~\bibnamefont
  {Bricard}}, \bibinfo {author} {\bibfnamefont {J.-B.}\ \bibnamefont
  {Caussin}}, \bibinfo {author} {\bibfnamefont {D.}~\bibnamefont {Das}},
  \bibinfo {author} {\bibfnamefont {C.}~\bibnamefont {Savoie}}, \bibinfo
  {author} {\bibfnamefont {V.}~\bibnamefont {Chikkadi}}, \bibinfo {author}
  {\bibfnamefont {K.}~\bibnamefont {Shitara}}, \bibinfo {author} {\bibfnamefont
  {O.}~\bibnamefont {Chepizhko}}, \bibinfo {author} {\bibfnamefont
  {F.}~\bibnamefont {Peruani}}, \bibinfo {author} {\bibfnamefont
  {D.}~\bibnamefont {Saintillan}},\ and\ \bibinfo {author} {\bibfnamefont
  {D.}~\bibnamefont {Bartolo}},\ }\href {https://doi.org/10.1038/ncomms8470}
  {\bibfield  {journal} {\bibinfo  {journal} {Nat. Commun.}\ }\textbf {\bibinfo
  {volume} {6}},\ \bibinfo {pages} {7470} (\bibinfo {year} {2015})}\BibitemShut
  {NoStop}%
\bibitem [{\citenamefont {Kawai}\ \emph {et~al.}(2020)\citenamefont {Kawai},
  \citenamefont {Matsunaga}, \citenamefont {Meng}, \citenamefont {Yeomans},\
  and\ \citenamefont {Golestanian}}]{kawai2020degenerate}%
  \BibitemOpen
  \bibfield  {author} {\bibinfo {author} {\bibfnamefont {T.}~\bibnamefont
  {Kawai}}, \bibinfo {author} {\bibfnamefont {D.}~\bibnamefont {Matsunaga}},
  \bibinfo {author} {\bibfnamefont {F.}~\bibnamefont {Meng}}, \bibinfo {author}
  {\bibfnamefont {J.~M.}\ \bibnamefont {Yeomans}},\ and\ \bibinfo {author}
  {\bibfnamefont {R.}~\bibnamefont {Golestanian}},\ }\href
  {https://doi.org/10.1039/D0SM00454E} {\bibfield  {journal} {\bibinfo
  {journal} {Soft Matter}\ }\textbf {\bibinfo {volume} {16}},\ \bibinfo {pages}
  {6484} (\bibinfo {year} {2020})}\BibitemShut {NoStop}%
\bibitem [{\citenamefont {Junot}\ \emph {et~al.}(2021)\citenamefont {Junot},
  \citenamefont {Cebers},\ and\ \citenamefont {Tierno}}]{junot2021collective}%
  \BibitemOpen
  \bibfield  {author} {\bibinfo {author} {\bibfnamefont {G.}~\bibnamefont
  {Junot}}, \bibinfo {author} {\bibfnamefont {A.}~\bibnamefont {Cebers}},\ and\
  \bibinfo {author} {\bibfnamefont {P.}~\bibnamefont {Tierno}},\ }\href
  {https://doi.org/10.1039/D1SM00653C} {\bibfield  {journal} {\bibinfo
  {journal} {Soft Matter}\ }\textbf {\bibinfo {volume} {17}},\ \bibinfo {pages}
  {8605} (\bibinfo {year} {2021})}\BibitemShut {NoStop}%
\bibitem [{\citenamefont {Yu}\ \emph {et~al.}(2018)\citenamefont {Yu},
  \citenamefont {Yang},\ and\ \citenamefont {Zhang}}]{yu2018pattern}%
  \BibitemOpen
  \bibfield  {author} {\bibinfo {author} {\bibfnamefont {J.}~\bibnamefont
  {Yu}}, \bibinfo {author} {\bibfnamefont {L.}~\bibnamefont {Yang}},\ and\
  \bibinfo {author} {\bibfnamefont {L.}~\bibnamefont {Zhang}},\ }\href
  {https://doi.org/10.1177/0278364918784366} {\bibfield  {journal} {\bibinfo
  {journal} {Int. J. Robotics Res.}\ }\textbf {\bibinfo {volume} {37}},\
  \bibinfo {pages} {912} (\bibinfo {year} {2018})}\BibitemShut {NoStop}%
\bibitem [{\citenamefont {Xie}\ \emph {et~al.}(2019)\citenamefont {Xie},
  \citenamefont {Sun}, \citenamefont {Fan}, \citenamefont {Lin}, \citenamefont
  {Chen}, \citenamefont {Wang}, \citenamefont {Dong},\ and\ \citenamefont
  {He}}]{xie2019reconfigurable}%
  \BibitemOpen
  \bibfield  {author} {\bibinfo {author} {\bibfnamefont {H.}~\bibnamefont
  {Xie}}, \bibinfo {author} {\bibfnamefont {M.}~\bibnamefont {Sun}}, \bibinfo
  {author} {\bibfnamefont {X.}~\bibnamefont {Fan}}, \bibinfo {author}
  {\bibfnamefont {Z.}~\bibnamefont {Lin}}, \bibinfo {author} {\bibfnamefont
  {W.}~\bibnamefont {Chen}}, \bibinfo {author} {\bibfnamefont {L.}~\bibnamefont
  {Wang}}, \bibinfo {author} {\bibfnamefont {L.}~\bibnamefont {Dong}},\ and\
  \bibinfo {author} {\bibfnamefont {Q.}~\bibnamefont {He}},\ }\bibfield
  {journal} {\bibinfo  {journal} {Sci. Robot.}\ }\textbf {\bibinfo {volume}
  {4}},\ \href {https://doi.org/10.1126/scirobotics.aav8006}
  {10.1126/scirobotics.aav8006} (\bibinfo {year} {2019})\BibitemShut {NoStop}%
\bibitem [{\citenamefont {Dorigo}\ \emph {et~al.}(2021)\citenamefont {Dorigo},
  \citenamefont {Theraulaz},\ and\ \citenamefont {Trianni}}]{dorigo2021swarm}%
  \BibitemOpen
  \bibfield  {author} {\bibinfo {author} {\bibfnamefont {M.}~\bibnamefont
  {Dorigo}}, \bibinfo {author} {\bibfnamefont {G.}~\bibnamefont {Theraulaz}},\
  and\ \bibinfo {author} {\bibfnamefont {V.}~\bibnamefont {Trianni}},\ }\href
  {https://doi.org/10.1109/JPROC.2021.3072740} {\bibfield  {journal} {\bibinfo
  {journal} {Proc. IEEE}\ }\textbf {\bibinfo {volume} {109}},\ \bibinfo {pages}
  {1152} (\bibinfo {year} {2021})}\BibitemShut {NoStop}%
\bibitem [{\citenamefont {Hussein}\ \emph {et~al.}(2023)\citenamefont
  {Hussein}, \citenamefont {Damdam}, \citenamefont {Ren}, \citenamefont
  {Obeid~Charrouf}, \citenamefont {Challita}, \citenamefont {Zwain},\ and\
  \citenamefont {Fariborzi}}]{hussein2023actuation}%
  \BibitemOpen
  \bibfield  {author} {\bibinfo {author} {\bibfnamefont {H.}~\bibnamefont
  {Hussein}}, \bibinfo {author} {\bibfnamefont {A.}~\bibnamefont {Damdam}},
  \bibinfo {author} {\bibfnamefont {L.}~\bibnamefont {Ren}}, \bibinfo {author}
  {\bibfnamefont {Y.}~\bibnamefont {Obeid~Charrouf}}, \bibinfo {author}
  {\bibfnamefont {J.}~\bibnamefont {Challita}}, \bibinfo {author}
  {\bibfnamefont {M.}~\bibnamefont {Zwain}},\ and\ \bibinfo {author}
  {\bibfnamefont {H.}~\bibnamefont {Fariborzi}},\ }\href
  {https://doi.org/10.1002/aisy.202300168} {\bibfield  {journal} {\bibinfo
  {journal} {Adv. Intell. Syst.}\ }\textbf {\bibinfo {volume} {n/a}},\ \bibinfo
  {pages} {2300168} (\bibinfo {year} {2023})}\BibitemShut {NoStop}%
\bibitem [{\citenamefont {Yu}\ \emph {et~al.}(2023)\citenamefont {Yu},
  \citenamefont {Li}, \citenamefont {Mou}, \citenamefont {Yu}, \citenamefont
  {Zhang}, \citenamefont {Yang}, \citenamefont {Zhao}, \citenamefont {Ma},
  \citenamefont {Luo}, \citenamefont {Li},\ and\ \citenamefont
  {Guan}}]{yu2023swarming}%
  \BibitemOpen
  \bibfield  {author} {\bibinfo {author} {\bibfnamefont {Z.}~\bibnamefont
  {Yu}}, \bibinfo {author} {\bibfnamefont {L.}~\bibnamefont {Li}}, \bibinfo
  {author} {\bibfnamefont {F.}~\bibnamefont {Mou}}, \bibinfo {author}
  {\bibfnamefont {S.}~\bibnamefont {Yu}}, \bibinfo {author} {\bibfnamefont
  {D.}~\bibnamefont {Zhang}}, \bibinfo {author} {\bibfnamefont
  {M.}~\bibnamefont {Yang}}, \bibinfo {author} {\bibfnamefont {Q.}~\bibnamefont
  {Zhao}}, \bibinfo {author} {\bibfnamefont {H.}~\bibnamefont {Ma}}, \bibinfo
  {author} {\bibfnamefont {W.}~\bibnamefont {Luo}}, \bibinfo {author}
  {\bibfnamefont {T.}~\bibnamefont {Li}},\ and\ \bibinfo {author}
  {\bibfnamefont {J.}~\bibnamefont {Guan}},\ }\href
  {https://doi.org/10.1002/inf2.12464} {\bibfield  {journal} {\bibinfo
  {journal} {InfoMat}\ }\textbf {\bibinfo {volume} {n/a}},\ \bibinfo {pages}
  {e12464} (\bibinfo {year} {2023})}\BibitemShut {NoStop}%
\bibitem [{\citenamefont {Hou}\ \emph {et~al.}(2023)\citenamefont {Hou},
  \citenamefont {Wang}, \citenamefont {Fu}, \citenamefont {Wang}, \citenamefont
  {Yu}, \citenamefont {Zhang}, \citenamefont {Huang}, \citenamefont {Sun},\
  and\ \citenamefont {Fukuda}}]{hou2023review}%
  \BibitemOpen
  \bibfield  {author} {\bibinfo {author} {\bibfnamefont {Y.}~\bibnamefont
  {Hou}}, \bibinfo {author} {\bibfnamefont {H.}~\bibnamefont {Wang}}, \bibinfo
  {author} {\bibfnamefont {R.}~\bibnamefont {Fu}}, \bibinfo {author}
  {\bibfnamefont {X.}~\bibnamefont {Wang}}, \bibinfo {author} {\bibfnamefont
  {J.}~\bibnamefont {Yu}}, \bibinfo {author} {\bibfnamefont {S.}~\bibnamefont
  {Zhang}}, \bibinfo {author} {\bibfnamefont {Q.}~\bibnamefont {Huang}},
  \bibinfo {author} {\bibfnamefont {Y.}~\bibnamefont {Sun}},\ and\ \bibinfo
  {author} {\bibfnamefont {T.}~\bibnamefont {Fukuda}},\ }\href
  {https://doi.org/10.1039/D2LC00573E} {\bibfield  {journal} {\bibinfo
  {journal} {Lab Chip}\ }\textbf {\bibinfo {volume} {23}},\ \bibinfo {pages}
  {848} (\bibinfo {year} {2023})}\BibitemShut {NoStop}%
\bibitem [{\citenamefont {{G{\'o}mez-Nava}}\ \emph {et~al.}(2022)\citenamefont
  {{G{\'o}mez-Nava}}, \citenamefont {Bon},\ and\ \citenamefont
  {Peruani}}]{gomez-nava2022intermittent}%
  \BibitemOpen
  \bibfield  {author} {\bibinfo {author} {\bibfnamefont {L.}~\bibnamefont
  {{G{\'o}mez-Nava}}}, \bibinfo {author} {\bibfnamefont {R.}~\bibnamefont
  {Bon}},\ and\ \bibinfo {author} {\bibfnamefont {F.}~\bibnamefont {Peruani}},\
  }\href {https://doi.org/10.1038/s41567-022-01769-8} {\bibfield  {journal}
  {\bibinfo  {journal} {Nat. Phys.}\ }\textbf {\bibinfo {volume} {18}},\
  \bibinfo {pages} {1} (\bibinfo {year} {2022})}\BibitemShut {NoStop}%
\bibitem [{\citenamefont {Wang}\ \emph {et~al.}(2022)\citenamefont {Wang},
  \citenamefont {Escobedo}, \citenamefont {Sanchez}, \citenamefont {Sire},
  \citenamefont {Han},\ and\ \citenamefont {Theraulaz}}]{wang2022impact}%
  \BibitemOpen
  \bibfield  {author} {\bibinfo {author} {\bibfnamefont {W.}~\bibnamefont
  {Wang}}, \bibinfo {author} {\bibfnamefont {R.}~\bibnamefont {Escobedo}},
  \bibinfo {author} {\bibfnamefont {S.}~\bibnamefont {Sanchez}}, \bibinfo
  {author} {\bibfnamefont {C.}~\bibnamefont {Sire}}, \bibinfo {author}
  {\bibfnamefont {Z.}~\bibnamefont {Han}},\ and\ \bibinfo {author}
  {\bibfnamefont {G.}~\bibnamefont {Theraulaz}},\ }\href
  {https://doi.org/10.1371/journal.pcbi.1009437} {\bibfield  {journal}
  {\bibinfo  {journal} {PLoS Comput. Biol.}\ }\textbf {\bibinfo {volume}
  {18}},\ \bibinfo {pages} {e1009437} (\bibinfo {year} {2022})}\BibitemShut
  {NoStop}%
\bibitem [{\citenamefont {Gompper}\ \emph {et~al.}(2020)\citenamefont {Gompper}
  \emph {et~al.}}]{gompper2020}%
  \BibitemOpen
  \bibfield  {author} {\bibinfo {author} {\bibfnamefont {G.}~\bibnamefont
  {Gompper}} \emph {et~al.},\ }\href {https://doi.org/10.1088/1361-648X/ab6348}
  {\bibfield  {journal} {\bibinfo  {journal} {J. Phys.: Condens. Matter}\
  }\textbf {\bibinfo {volume} {32}},\ \bibinfo {pages} {193001} (\bibinfo
  {year} {2020})}\BibitemShut {NoStop}%
\bibitem [{\citenamefont {Abaurrea~Velasco}\ \emph {et~al.}(2018)\citenamefont
  {Abaurrea~Velasco}, \citenamefont {Abkenar}, \citenamefont {Gompper},\ and\
  \citenamefont {Auth}}]{abaurreavelasco2018collective}%
  \BibitemOpen
  \bibfield  {author} {\bibinfo {author} {\bibfnamefont {C.}~\bibnamefont
  {Abaurrea~Velasco}}, \bibinfo {author} {\bibfnamefont {M.}~\bibnamefont
  {Abkenar}}, \bibinfo {author} {\bibfnamefont {G.}~\bibnamefont {Gompper}},\
  and\ \bibinfo {author} {\bibfnamefont {T.}~\bibnamefont {Auth}},\ }\href
  {https://doi.org/10.1103/PhysRevE.98.022605} {\bibfield  {journal} {\bibinfo
  {journal} {Phys. Rev. E}\ }\textbf {\bibinfo {volume} {98}},\ \bibinfo
  {pages} {022605} (\bibinfo {year} {2018})}\BibitemShut {NoStop}%
\bibitem [{\citenamefont {Wang}\ \emph {et~al.}(2017)\citenamefont {Wang},
  \citenamefont {Giltinan}, \citenamefont {Zakharchenko},\ and\ \citenamefont
  {Sitti}}]{wang2017dynamic}%
  \BibitemOpen
  \bibfield  {author} {\bibinfo {author} {\bibfnamefont {W.}~\bibnamefont
  {Wang}}, \bibinfo {author} {\bibfnamefont {J.}~\bibnamefont {Giltinan}},
  \bibinfo {author} {\bibfnamefont {S.}~\bibnamefont {Zakharchenko}},\ and\
  \bibinfo {author} {\bibfnamefont {M.}~\bibnamefont {Sitti}},\ }\href
  {https://doi.org/10.1126/sciadv.1602522} {\bibfield  {journal} {\bibinfo
  {journal} {Sci. Adv.}\ }\textbf {\bibinfo {volume} {3}},\ \bibinfo {pages}
  {e1602522} (\bibinfo {year} {2017})}\BibitemShut {NoStop}%
\bibitem [{\citenamefont {Nava}\ \emph {et~al.}(2020)\citenamefont {Nava},
  \citenamefont {Gro{\ss}mann}, \citenamefont {Hintsche}, \citenamefont
  {Beta},\ and\ \citenamefont {Peruani}}]{nava2020novel}%
  \BibitemOpen
  \bibfield  {author} {\bibinfo {author} {\bibfnamefont {L.~G.}\ \bibnamefont
  {Nava}}, \bibinfo {author} {\bibfnamefont {R.}~\bibnamefont {Gro{\ss}mann}},
  \bibinfo {author} {\bibfnamefont {M.}~\bibnamefont {Hintsche}}, \bibinfo
  {author} {\bibfnamefont {C.}~\bibnamefont {Beta}},\ and\ \bibinfo {author}
  {\bibfnamefont {F.}~\bibnamefont {Peruani}},\ }\href
  {https://doi.org/10.1209/0295-5075/130/68002} {\bibfield  {journal} {\bibinfo
   {journal} {EPL}\ }\textbf {\bibinfo {volume} {130}},\ \bibinfo {pages}
  {68002} (\bibinfo {year} {2020})}\BibitemShut {NoStop}%
\bibitem [{\citenamefont {Ceron}\ \emph {et~al.}(2023)\citenamefont {Ceron},
  \citenamefont {Gardi}, \citenamefont {Petersen},\ and\ \citenamefont
  {Sitti}}]{ceron2023programmable}%
  \BibitemOpen
  \bibfield  {author} {\bibinfo {author} {\bibfnamefont {S.}~\bibnamefont
  {Ceron}}, \bibinfo {author} {\bibfnamefont {G.}~\bibnamefont {Gardi}},
  \bibinfo {author} {\bibfnamefont {K.}~\bibnamefont {Petersen}},\ and\
  \bibinfo {author} {\bibfnamefont {M.}~\bibnamefont {Sitti}},\ }\href
  {https://doi.org/10.1073/pnas.2221913120} {\bibfield  {journal} {\bibinfo
  {journal} {Proc. Natl. Acad. Sci.}\ }\textbf {\bibinfo {volume} {120}},\
  \bibinfo {pages} {e2221913120} (\bibinfo {year} {2023})}\BibitemShut
  {NoStop}%
\bibitem [{\citenamefont {Stevens}\ and\ \citenamefont
  {Greenberg}(1997)}]{stevens1997quorum}%
  \BibitemOpen
  \bibfield  {author} {\bibinfo {author} {\bibfnamefont {A.~M.}\ \bibnamefont
  {Stevens}}\ and\ \bibinfo {author} {\bibfnamefont {E.~P.}\ \bibnamefont
  {Greenberg}},\ }\href {https://doi.org/10.1128/jb.179.2.557-562.1997}
  {\bibfield  {journal} {\bibinfo  {journal} {J. Bacteriol.}\ }\textbf
  {\bibinfo {volume} {179}},\ \bibinfo {pages} {557} (\bibinfo {year}
  {1997})}\BibitemShut {NoStop}%
\bibitem [{\citenamefont {B{\"a}uerle}\ \emph {et~al.}(2018)\citenamefont
  {B{\"a}uerle}, \citenamefont {Fischer}, \citenamefont {Speck},\ and\
  \citenamefont {Bechinger}}]{bauerle2018selforganization}%
  \BibitemOpen
  \bibfield  {author} {\bibinfo {author} {\bibfnamefont {T.}~\bibnamefont
  {B{\"a}uerle}}, \bibinfo {author} {\bibfnamefont {A.}~\bibnamefont
  {Fischer}}, \bibinfo {author} {\bibfnamefont {T.}~\bibnamefont {Speck}},\
  and\ \bibinfo {author} {\bibfnamefont {C.}~\bibnamefont {Bechinger}},\ }\href
  {https://doi.org/10.1038/s41467-018-05675-7} {\bibfield  {journal} {\bibinfo
  {journal} {Nat. Commun.}\ }\textbf {\bibinfo {volume} {9}},\ \bibinfo {pages}
  {3232} (\bibinfo {year} {2018})}\BibitemShut {NoStop}%
\bibitem [{\citenamefont {Barberis}\ and\ \citenamefont
  {Peruani}(2016)}]{barberis2016largescale}%
  \BibitemOpen
  \bibfield  {author} {\bibinfo {author} {\bibfnamefont {L.}~\bibnamefont
  {Barberis}}\ and\ \bibinfo {author} {\bibfnamefont {F.}~\bibnamefont
  {Peruani}},\ }\href {https://doi.org/10.1103/PhysRevLett.117.248001}
  {\bibfield  {journal} {\bibinfo  {journal} {Phys. Rev. Lett.}\ }\textbf
  {\bibinfo {volume} {117}},\ \bibinfo {pages} {248001} (\bibinfo {year}
  {2016})}\BibitemShut {NoStop}%
\bibitem [{\citenamefont {Costanzo}(2019)}]{costanzo2019millinginduction}%
  \BibitemOpen
  \bibfield  {author} {\bibinfo {author} {\bibfnamefont {A.}~\bibnamefont
  {Costanzo}},\ }\href {https://doi.org/10.1209/0295-5075/125/20008} {\bibfield
   {journal} {\bibinfo  {journal} {EPL}\ }\textbf {\bibinfo {volume} {125}},\
  \bibinfo {pages} {20008} (\bibinfo {year} {2019})}\BibitemShut {NoStop}%
\bibitem [{\citenamefont {Negi}\ \emph {et~al.}(2022)\citenamefont {Negi},
  \citenamefont {Winkler},\ and\ \citenamefont {Gompper}}]{negi2022emergent}%
  \BibitemOpen
  \bibfield  {author} {\bibinfo {author} {\bibfnamefont {R.~S.}\ \bibnamefont
  {Negi}}, \bibinfo {author} {\bibfnamefont {R.~G.}\ \bibnamefont {Winkler}},\
  and\ \bibinfo {author} {\bibfnamefont {G.}~\bibnamefont {Gompper}},\ }\href
  {https://doi.org/10.1039/D2SM00736C} {\bibfield  {journal} {\bibinfo
  {journal} {Soft Matter}\ }\textbf {\bibinfo {volume} {18}},\ \bibinfo {pages}
  {6167} (\bibinfo {year} {2022})}\BibitemShut {NoStop}%
\bibitem [{\citenamefont {Negi}\ \emph {et~al.}(2024)\citenamefont {Negi},
  \citenamefont {Winkler},\ and\ \citenamefont {Gompper}}]{negi2024collective}%
  \BibitemOpen
  \bibfield  {author} {\bibinfo {author} {\bibfnamefont {R.~S.}\ \bibnamefont
  {Negi}}, \bibinfo {author} {\bibfnamefont {R.~G.}\ \bibnamefont {Winkler}},\
  and\ \bibinfo {author} {\bibfnamefont {G.}~\bibnamefont {Gompper}},\ }\href
  {https://doi.org/10.1103/PhysRevResearch.6.013118} {\bibfield  {journal}
  {\bibinfo  {journal} {Phys. Rev. Res.}\ }\textbf {\bibinfo {volume} {6}},\
  \bibinfo {pages} {013118} (\bibinfo {year} {2024})}\BibitemShut {NoStop}%
\bibitem [{\citenamefont {B{\"a}uerle}\ \emph {et~al.}(2020)\citenamefont
  {B{\"a}uerle}, \citenamefont {L{\"o}ffler},\ and\ \citenamefont
  {Bechinger}}]{bauerle2020formation}%
  \BibitemOpen
  \bibfield  {author} {\bibinfo {author} {\bibfnamefont {T.}~\bibnamefont
  {B{\"a}uerle}}, \bibinfo {author} {\bibfnamefont {R.~C.}\ \bibnamefont
  {L{\"o}ffler}},\ and\ \bibinfo {author} {\bibfnamefont {C.}~\bibnamefont
  {Bechinger}},\ }\href {https://doi.org/10.1038/s41467-020-16161-4} {\bibfield
   {journal} {\bibinfo  {journal} {Nat. Commun.}\ }\textbf {\bibinfo {volume}
  {11}},\ \bibinfo {pages} {2547} (\bibinfo {year} {2020})}\BibitemShut
  {NoStop}%
\bibitem [{\citenamefont {Wysocki}\ \emph {et~al.}(2016)\citenamefont
  {Wysocki}, \citenamefont {Winkler},\ and\ \citenamefont
  {Gompper}}]{wysocki2016propagating}%
  \BibitemOpen
  \bibfield  {author} {\bibinfo {author} {\bibfnamefont {A.}~\bibnamefont
  {Wysocki}}, \bibinfo {author} {\bibfnamefont {R.~G.}\ \bibnamefont
  {Winkler}},\ and\ \bibinfo {author} {\bibfnamefont {G.}~\bibnamefont
  {Gompper}},\ }\href {https://doi.org/10.1088/1367-2630/aa529d} {\bibfield
  {journal} {\bibinfo  {journal} {New J. Phys.}\ }\textbf {\bibinfo {volume}
  {18}},\ \bibinfo {pages} {123030} (\bibinfo {year} {2016})}\BibitemShut
  {NoStop}%
\bibitem [{\citenamefont {{Agudo-Canalejo}}\ and\ \citenamefont
  {Golestanian}(2019)}]{agudo-canalejo2019active}%
  \BibitemOpen
  \bibfield  {author} {\bibinfo {author} {\bibfnamefont {J.}~\bibnamefont
  {{Agudo-Canalejo}}}\ and\ \bibinfo {author} {\bibfnamefont {R.}~\bibnamefont
  {Golestanian}},\ }\href {https://doi.org/10.1103/PhysRevLett.123.018101}
  {\bibfield  {journal} {\bibinfo  {journal} {Phys. Rev. Lett.}\ }\textbf
  {\bibinfo {volume} {123}},\ \bibinfo {pages} {018101} (\bibinfo {year}
  {2019})}\BibitemShut {NoStop}%
\bibitem [{\citenamefont {Chen}\ and\ \citenamefont
  {Bechinger}(2022)}]{chen2022collective}%
  \BibitemOpen
  \bibfield  {author} {\bibinfo {author} {\bibfnamefont {C.-J.}\ \bibnamefont
  {Chen}}\ and\ \bibinfo {author} {\bibfnamefont {C.}~\bibnamefont
  {Bechinger}},\ }\href {https://doi.org/10.1088/1367-2630/ac5374} {\bibfield
  {journal} {\bibinfo  {journal} {New J. Phys.}\ }\textbf {\bibinfo {volume}
  {24}},\ \bibinfo {pages} {033001} (\bibinfo {year} {2022})}\BibitemShut
  {NoStop}%
\bibitem [{\citenamefont {B{\"a}r}\ \emph {et~al.}(2020)\citenamefont
  {B{\"a}r}, \citenamefont {Gro{\ss}mann}, \citenamefont {Heidenreich},\ and\
  \citenamefont {Peruani}}]{bar2020selfpropelled}%
  \BibitemOpen
  \bibfield  {author} {\bibinfo {author} {\bibfnamefont {M.}~\bibnamefont
  {B{\"a}r}}, \bibinfo {author} {\bibfnamefont {R.}~\bibnamefont
  {Gro{\ss}mann}}, \bibinfo {author} {\bibfnamefont {S.}~\bibnamefont
  {Heidenreich}},\ and\ \bibinfo {author} {\bibfnamefont {F.}~\bibnamefont
  {Peruani}},\ }\href
  {https://doi.org/10.1146/annurev-conmatphys-031119-050611} {\bibfield
  {journal} {\bibinfo  {journal} {Annu. Rev. Condens. Matter Phys.}\ }\textbf
  {\bibinfo {volume} {11}},\ \bibinfo {pages} {441} (\bibinfo {year}
  {2020})}\BibitemShut {NoStop}%
\bibitem [{\citenamefont {Caprini}\ and\ \citenamefont
  {L{\"o}wen}(2023)}]{caprini2023flocking}%
  \BibitemOpen
  \bibfield  {author} {\bibinfo {author} {\bibfnamefont {L.}~\bibnamefont
  {Caprini}}\ and\ \bibinfo {author} {\bibfnamefont {H.}~\bibnamefont
  {L{\"o}wen}},\ }\href {https://doi.org/10.1103/PhysRevLett.130.148202}
  {\bibfield  {journal} {\bibinfo  {journal} {Phys. Rev. Lett.}\ }\textbf
  {\bibinfo {volume} {130}},\ \bibinfo {pages} {148202} (\bibinfo {year}
  {2023})}\BibitemShut {NoStop}%
\bibitem [{\citenamefont {Lavergne}\ \emph {et~al.}(2019)\citenamefont
  {Lavergne}, \citenamefont {Wendehenne}, \citenamefont {B{\"a}uerle},\ and\
  \citenamefont {Bechinger}}]{lavergne2019group}%
  \BibitemOpen
  \bibfield  {author} {\bibinfo {author} {\bibfnamefont {F.~A.}\ \bibnamefont
  {Lavergne}}, \bibinfo {author} {\bibfnamefont {H.}~\bibnamefont
  {Wendehenne}}, \bibinfo {author} {\bibfnamefont {T.}~\bibnamefont
  {B{\"a}uerle}},\ and\ \bibinfo {author} {\bibfnamefont {C.}~\bibnamefont
  {Bechinger}},\ }\href {https://doi.org/10.1126/science.aau5347} {\bibfield
  {journal} {\bibinfo  {journal} {Science}\ }\textbf {\bibinfo {volume}
  {364}},\ \bibinfo {pages} {70} (\bibinfo {year} {2019})}\BibitemShut
  {NoStop}%
\bibitem [{\citenamefont {Saavedra}\ \emph {et~al.}(2024)\citenamefont
  {Saavedra}, \citenamefont {Gompper},\ and\ \citenamefont
  {Ripoll}}]{saavedra2024swirling}%
  \BibitemOpen
  \bibfield  {author} {\bibinfo {author} {\bibfnamefont {R.}~\bibnamefont
  {Saavedra}}, \bibinfo {author} {\bibfnamefont {G.}~\bibnamefont {Gompper}},\
  and\ \bibinfo {author} {\bibfnamefont {M.}~\bibnamefont {Ripoll}},\ }\href
  {https://doi.org/10.1103/PhysRevLett.132.268301} {\bibfield  {journal}
  {\bibinfo  {journal} {Phys. Rev. Lett.}\ }\textbf {\bibinfo {volume} {132}},\
  \bibinfo {pages} {268301} (\bibinfo {year} {2024})}\BibitemShut {NoStop}%
\bibitem [{sm()}]{sm}%
  \BibitemOpen
  \href@noop {} {}\bibinfo {note} {See Supplementary Materials in Ref.xxx,
  .}\BibitemShut {Stop}%
\bibitem [{\citenamefont {Moran}(1950)}]{moran}%
  \BibitemOpen
  \bibfield  {author} {\bibinfo {author} {\bibfnamefont {P.}~\bibnamefont
  {Moran}},\ }\href {https://doi.org/https://doi.org/10.1093/biomet/37.1-2.17}
  {\bibfield  {journal} {\bibinfo  {journal} {Biometrika}\ }\textbf {\bibinfo
  {volume} {37}},\ \bibinfo {pages} {17} (\bibinfo {year} {1950})}\BibitemShut
  {NoStop}%
\bibitem [{\citenamefont {Dirk}\ \emph {et~al.}(2023)\citenamefont {Dirk},
  \citenamefont {Fischer}, \citenamefont {Schardt}, \citenamefont
  {Ankenbrand},\ and\ \citenamefont {Fischer}}]{fischer23}%
  \BibitemOpen
  \bibfield  {author} {\bibinfo {author} {\bibfnamefont {R.}~\bibnamefont
  {Dirk}}, \bibinfo {author} {\bibfnamefont {J.~L.}\ \bibnamefont {Fischer}},
  \bibinfo {author} {\bibfnamefont {S.}~\bibnamefont {Schardt}}, \bibinfo
  {author} {\bibfnamefont {M.~J.}\ \bibnamefont {Ankenbrand}},\ and\ \bibinfo
  {author} {\bibfnamefont {S.~C.}\ \bibnamefont {Fischer}},\ }\href
  {https://doi.org/https://doi.org/10.1371/journal.pcbi.1011582} {\bibfield
  {journal} {\bibinfo  {journal} {PLoS Comput. Biol.}\ }\textbf {\bibinfo
  {volume} {19}},\ \bibinfo {pages} {e1011582} (\bibinfo {year}
  {2023})}\BibitemShut {NoStop}%
\bibitem [{\citenamefont {Centre}(2018)}]{jureca2018}%
  \BibitemOpen
  \bibfield  {author} {\bibinfo {author} {\bibfnamefont {J.~S.}\ \bibnamefont
  {Centre}},\ }\href {https://doi.org/10.17815/jlsrf-7-182} {\bibfield
  {journal} {\bibinfo  {journal} {Journal of large-scale research facilities}\
  }\textbf {\bibinfo {volume} {7}},\ \bibinfo {pages} {A182} (\bibinfo {year}
  {2018})}\BibitemShut {NoStop}%
\end{thebibliography}

%

\end{document}